\documentclass[aps,prl,twocolumn,superscriptaddress]{revtex4}
\usepackage{epsfig,graphicx,times}
\def\lsim{\hbox{\lower .8ex\hbox{$\, \buildrel < \over \sim\,$}}}
\def\gsim{\hbox{\lower .8ex\hbox{$\, \buildrel > \over \sim\,$}}}

\newcommand{\Veff}{V_{\rm eff}}

\newcommand{\Vsc}{V_{\rm sc}}

\newcommand{\Heff}{\hat H_{\rm eff}}

\newcommand{\Etf}{{\cal E}_{\rm TF}}
\newcommand{\Eop}{{\cal E}_{\rm 1p}}
\newcommand{\Eri}{{\cal E}_{\rm ri}}

\newcommand{\Egr}{E_{\rm gr}}

\newcommand{\br}{{\bf r}}

\newcommand{\bp}{{\bf p}}

\newcommand{\e}{\epsilon}

\newcommand{\s}{\sigma}

\begin{document}

\title{Quantum  dot  ground  state  energies and  spin  polarizations:
soft versus hard chaos}

\author{Denis Ullmo}
\affiliation{Laboratoire~de~Physique~Th\'eorique et Mod\`eles~Statistiques
(LPTMS), 91405 Orsay Cedex,~France}
\affiliation{Department of Physics, Duke University, Durham, North
    Carolina 27708-0305, USA}
\author{Tatsuro Nagano}
\affiliation{Department of
Physics, Washington State University, Pullman, WA 99164-2814, USA}
\author{Steven   Tomsovic}
\affiliation{Department of
Physics, Washington State University, Pullman, WA 99164-2814, USA}

\date{\today}
\begin{abstract}

We  consider how the nature of the dynamics affects ground state
properties of ballistic quantum dots.  We find that ``mesoscopic Stoner
fluctuations'', that arise from the residual screened Coulomb interaction,
are very sensitive to the degree of chaos.  It leads to ground state
energies and spin-polarizations whose fluctuations strongly
increase as a system becomes less chaotic.  The crucial features are
illustrated with a model that depends on a parameter that tunes the
dynamics from nearly integrable to mostly chaotic.

\end{abstract}



\maketitle

Our  interest  in this  letter  lies  in  microstructures fabricated using
electrostatic gates or etching that pattern a two dimensional
electron gas in a semiconductor heterostructure, for example,
GaAs/AlGaAs.   Typically, the electronic  transport mean free path is
significantly larger than  the dimensions of the  device, and the
electrons   essentially    travel   ballistically   across   the
microstructure.  Their  motion is governed  by the shape of  a smooth,
self-consistent,  steep-walled,  confining  potential which  is  often
conceptualized as a quantum billiard.

For many physical properties, the simplifying assumption that the dots'
underlying classical dynamics are fully chaotic (hard chaos) has
provided a good description of the experimental
data~\cite{rodolfo,albert,gonzalo,marcus_dot}.  It has been used to
justify various hypotheses from the applicability of random matrix
theory  (RMT) and  random  plane wave  modeling  (RPW) to  statistical
assumptions           applied           within           semiclassical
mechanics~\cite{justify,RMT_modeling,ullmo_baranger_01}.  Indeed,
chaotic systems manifest a large variety of {\em universal} behaviors.
Furthermore, chaotic quantum dots are often qualitatively very similar to
diffusive ones provided the Thouless energy $E_{TH}$ is defined as $\hbar
v_F / L$, where $v_F$ is the  Fermi velocity and $L$ is  a typical
dimension of  the dot (as opposed  to $\hbar  D/L^2$ with  $D$  the
diffusion  constant).  Consequently, most techniques, developed much
earlier to study disordered metals (diagrammatic approaches~\cite{A&A},
nonlinear sigma model~\cite{efetov}) and applied to disordered quantum
dots~\cite{blanter,glazman,DIFF_modeling}, are applicable to ballistic
quantum dots.

Nevertheless, unlike billiards, there are no known smooth potentials
which are truly, fully chaotic.   Unless designed otherwise for a
specific purpose (such as measuring the weak localization
lineshape~\cite{weak_loc_linsehape}) an odd-shaped, smooth potential
generically exhibits {\it soft} chaos, i.e.~significant contributions of
{\em both} stable and unstable motion.  The general assumption of hard
chaos is unfounded.

As opposed to a genuine belief that the electrons' dynamics are strongly
chaotic, the implicit assumption is that for many properties the
distinctions between soft and hard chaos are more subtle than
spectacular.  This has been shown explicitly, for instance, for the
fluctuation properties of Coulomb blockade (CB) peak heights~\cite{caio},
or for their correlations~\cite{narimanov}.  In such circumstances, using
a chaotic model allows for simpler analytic derivations without
drastically altering the results.

Our purpose  here is to  demonstrate that even this  weaker assumption
may, in  some cases, be problematic;  and that for  some properties or
measurements  a  strong sensitivity  to  the  nature  of the  dynamics
arises.  Because  the distinction  between the chaotic  and integrable
limits becomes  most apparent at long (infinite)  times, confining the
electrons as  long as  practical within the  dot holds the  promise of
leading to  signatures of the  dynamics.  We shall  therefore consider
[zero temperature] ground states properties of well isolated dots, and
find that  ground state energies (whose second  differences are probed
by CB peak spacing  measurements), and spin polarizations are markedly
affected by the dynamics.

We proceed as follows.  First, we give a general discussion of why, in
principle, chaotic dots should be rather ``atypical'', at least as far
as ground state properties are concerned.   Secondly,
we consider in more detail a particular Hamiltonian model, namely a
time-reversal non-invariant coupled quartic oscillator system, and
show more quantitatively the relevance of the underlying dynamics.

In the following, we assume a Fermi-Landau liquid description of
the quantum dot~\cite{blanter,tatsuro1,glazman}, and therefore
that the ground state energy $\Egr[N]$ is the sum of three terms
\begin{equation} \label{eq:Egr}
     \Egr[N] = \Etf + \Eop + \Eri \ \ .
\end{equation}
$\Etf[N] = (eN)^2/2C$ is an electrostatic energy, $\Eop[N]$ is the
sum of the single particle energies (SPE) of $N$ {\em independent}
particles, and $\Eri$ is a residual interaction term.  Specifically
\begin{equation} \label{eq:E1p}
\Eop[N] = \sum_{i,\s} f_{i}^{\s} \e_i \; ,
\end{equation}
with $\e_i$ the SPEs corresponding to an effective potential $\Veff(\br)$
(that can be determined self-consistently within the electrostatic-like
approximation) and $f_{i}^{\s} =  0$ or $1$ is the occupation number of
orbital $i$ with spin $\s = \pm 1$, ($\sum_{i,\s} f_{i}^{\s} = N$).
Furthermore, denoting $\psi_i$ the eigenstate associated to $\e_i$,
\begin{eqnarray*}
&& \Eri  =  \frac{1}{2} \sum_{i,j,\s,\s'} f_{i}^{\s} f_{j}^{\s'}
    \int d\br d\br' \left| \psi_i(\br) \right|^2
            \Vsc(\br-\br') \left| \psi_{j}(\br') \right|^2  \\
   &&      -  \frac{1}{2} \sum_{i,j,\sigma}f_{i}^{\s} f_{j}^{\s}
          \int d\br d\br' \psi_i(\br) \psi_{j}^*(\br)
            \Vsc(\br-\br') \psi_{j}(\br') \psi_{i}^*(\br')
\end{eqnarray*}
is the direct-plus-exchange, first order perturbation contribution in
terms of the weak screened Coulomb interaction $\Vsc$.  For the
experimentally relevant electronic densities, $\Vsc$ is a short range
function, and we shall model it as
\begin{equation} \label{eq:Vsc}
\Vsc = \frac{\zeta}{\nu} \delta(\br-\br')
\end{equation}
with $\nu$ the total density  of states (including the spin degeneracy
factor $g_s=2$) and $\zeta \in  [0,1]$ a parameter that can be related
to the density of electrons  (i.e.~the parameter $r_s$ of the electron
gas) and  is in the range  $[0.5, 0.8]$ for  many experiments.  Within
this zero range  approximation, the residual interaction contributions
read
\begin{equation} \label{eq:Eri}
	\Eri = \zeta \frac{\Delta}{g_s}
	\sum_{i,j} f_{i}^{(+)} f_{j}^{(-)}  M_{ij}
\end{equation}
with $\Delta$ the (local) mean SPE spacing and
\begin{equation} \label{eq:Mij}
M_{ij} =  A \int d\br \left| \psi_i(\br) \right|^2 \left| \psi_{j}(\br)
\right|^2  \; .
\end{equation}
$A$ is the dot area.  The $M_{ij}$ are dimensionless and semiclassical
reasoning implies that their mean (for $i \neq j$) is unity.

Note the crucial point that in Eq.~(\ref{eq:Eri}), only electrons with
opposite spins actually interact.  Aligning two spins decreases the
residual interaction by a quantity of $O(\Delta)$.  There is thus a
competition between the SPE term, which favors the occupation of the
lowest orbitals, and the residual interaction term, which tends to align
the electron spins.  The relative strength of these two effects
is governed by the dimensionless parameter $\zeta$.  If
$\zeta > 2$, the ground state is completely polarized.  This is the well
known Stoner instability~\cite{stoner}, which for instance is responsible
for the ferromagnetic character of cobalt.

Here $\zeta$ is just less than one.  Although full polarization is
excluded, the proximity of the Stoner instability makes the ground state
spin very sensitive to the fluctuations of the $M_{ij}$ and $\e_i$, which
affects the fluctuation properties of the ground
state~\cite{blanter,ullmo_baranger_01,glazman,altshuler}.    This is
sometimes referred to as the ``mesoscopic Stoner fluctuation''.

Now consider the diagonal term
\begin{equation} \label{eq:Mii}
M_{ii} =  A \int d\br \left| \psi_i(\br) \right|^4  \; .
\end{equation}
$M_{ii}$ is the inverse participation ratio in position representation
of the state $\psi_i$, which measures its extent of localization.
Hard chaotic systems possess eigenfunctions that are the least
localized in the sense that their Wigner transforms uniformly cover the
energy surface~\cite{voros}.  However, mixed systems are known to display
various forms of {\em phase space localization}~\cite{physrep}.  This
arises by various mechanisms.  The most familiar one is associated
with the quantization of invariant tori.  It affects only a minority of
states, but localizes them very strongly.  Another mechanism discussed
in~\cite{physrep} is associated with the presence of partial transport
barriers in phase space, the presence of which should be quite typical
in mixed dynamical systems.  Such partial barriers, if they are effective,
should affect almost all eigenstates, but produce a lesser degree of
localization.

Similarly, it can be shown that the mean value of the off-diagonal terms
$M_{i,j\neq i}$ should be independent of the dynamics, and their
fluctuations are extremely small for chaotic systems (vanishing in the
semiclassical limit~\cite{ullmo_baranger_01}).  However, they would be
of $O(\Delta)$ if significant phase space localization is present.  We
therefore see that chaotic systems, rather than behaving {\em typically},
are rather the limiting class of systems for which the interaction is
the {\em least} effective.

To explore the affect of soft chaos on the ground state properties,
we introduce a specific model.  Let the effective Hamiltonian $\Heff$
resulting from the lowest order, electrostatic-like, self-consistent
calculation be ($\br = (x,y)$, $r = |\br|)$)
\begin{equation}
   \Heff = \frac{\left(\bp - \kappa \sqrt{a(\lambda)} x^2 {\br \over
       r} \right)^2}{2}
   +
   a(\lambda) \left( \frac{x^4}{b} + b y^4 + 2\lambda x^2 y ^2 \right)
   \; .
   \nonumber
\end{equation}
Weyl operator ordering is assumed for the quantized version.  The
parameter $b=\pi/4$ is introduced so that the system has the symmetry of
the rectangle instead of the square.  $a(\lambda)$ is a convenient scaling
factor chosen so that the mean number of states with energy smaller than
$E$ is given by $N(E) = E^{3/2}$, regardless of the choice of $\lambda$
or $\kappa$.  $\lambda$ is the coupling  between  the  oscillators.
Finally, the parameter $\kappa$ breaks time reversal invariance (TRI).
Note that for TRI systems, higher order terms in the ground state energy
expansion, in particular the Cooper series, should be taken into
account.  Therefore, $\kappa$ is chosen such that TRI is completely
broken.  The specific form of the TRI breaking term has been chosen not
to pertain to any particular physical realization, but to insure that the
phase space portrait of the dynamics does not depend on the energy.  By
choosing $\lambda$ and $\kappa$ appropriately, various regimes of dynamics
can be studied. In particular, we consider $(\lambda,\kappa) =
(+0.20,1.00)$ [nearly  integrable], $(-0.20,1.00)$ [mixed], and
$(-0.80,1.00)$ [mostly chaotic].  Due to the TRI breaking term somewhat
stabilizing the dynamics, the motion in the chaotic case is still not
quite fully chaotic.

The reflection symmetries lead to four irreducible representations,
which can be thought of as independent quantum dots with the same
dynamics.  We consider the four systems as an ensemble, which allows us
to decrease our statistical `error bars' in the calculations.  In fact,
we increase the size of the ensemble even more by allowing $\lambda$ to
vary $\pm 0.02$, which is enough to get nearly independent quantum
eigenproperties, but small enough that the structure of the dynamics is
essentially unchanged.   Each statistical measure calculated within a
given dynamical regime is thus the result of averaging over an ensemble
of 12 similar quantum dots.

For each parameter set, it is possible to compute for each symmetry
class $(e_x=\pm 1,  e_y = \pm 1)$ the eigenvalues $\e_i$ and
eigenvectors  $\psi_i$, from which the residual interaction terms
Eq.~(\ref{eq:Mij}) can be deduced (see~\cite{tatsuro_thesis} for the
numerical  details).   We first consider a few interesting
statistical properties of these quantities, and afterward see how the
spin distribution and ground state energy fluctuations
are affected.
\begin{figure}[t]
\includegraphics[width=3.3in]{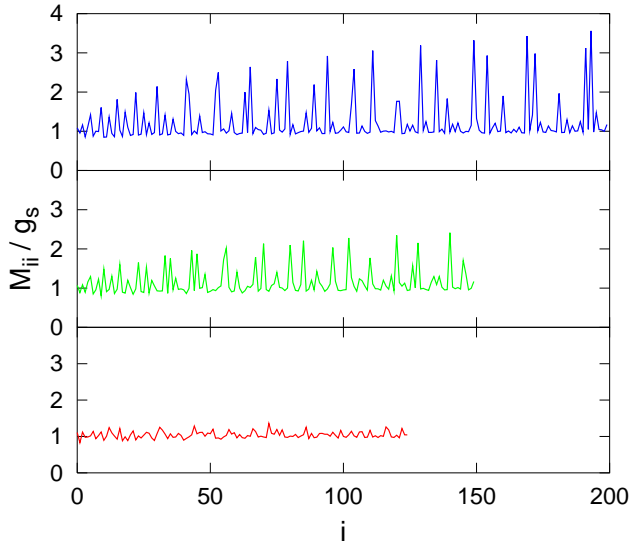}
\caption{Inverse  participation ratio as a function of the orbital index
for $(+,+)$.   From top down: $(\lambda,\kappa) = (+0.20,1.00)$
[nearly integrable], $(-0.20,1.00)$ [mixed], and $(-0.80,1.00)$ [mostly
chaotic].
\label{fig:miiScaled}}
\end{figure}

In Fig.~\ref{fig:miiScaled}, values of sets of diagonal terms $M_{ii}$
are represented for the symmetry class $(+,+)$ in the various dynamical
regimes.  The presence of very localized states in the nearly integrable
and mixed regimes is immediately apparent.  Curiously enough, it turns
out that for all dynamical regimes, the IPR  $M_{ii}$ is not very far from
two for most states, but the distribution has a very long tail if the
dynamics are not sufficiently chaotic.   Consequently, it is reasonable
to assume that the localization mechanism at work here is associated with
stable periodic orbits or tori.  A more detailed analysis of the
classical dynamics reveals that partial barriers are actually present
in $(\lambda,\kappa) =  (-0.20,1.00)$, but that their position with
respect to the symmetry lines of the system make them ineffective in
changing the statistics of the $M_{ii}$.  Furthermore, and as can be seen
in Table~\ref{table:Mij}, the off-diagonal terms $M_{ij}$ ($i \neq j$)
are also affected by eigenstate localization as their variance is
significantly larger if the state $i$ is localized.
\begin{table}[b]
\begin{tabular}{|c|ccc|}
\hline
$\lambda$        & \ +0.20 \ & \ -0.20 \ & \  -0.80 \ \\
$I^{\rm sup}$    &  2.0  & 1.8   & 1.2  \\
$I^{\rm inf}$    &  1.2  & 1.2   & 1.0  \\
\hline
$~~M_{ii}/g_s > I^{\rm sup}~~$    & ~~0.097~~~ & ~~0.108~~~ & ~~0.070~~~ \\
$M_{ii}  /g_s  < I^{\rm inf}$    & 0.024 & 0.023 & 0.009 \\
\hline
\end{tabular}
\caption{\label{table:Mij} Conditional variance of the interaction
terms $M_{ij}$ with $0 < |i-j| \le 10 $, $i\ge 51$.  Rows top down: i)
dynamical case, ii) superior limit, iii) inferior limit, iv) conditional
variance with localized orbitals, and v) conditional variance
with delocalized orbitals.}
\end{table}

\begin{figure}
\includegraphics[width=2.8in]{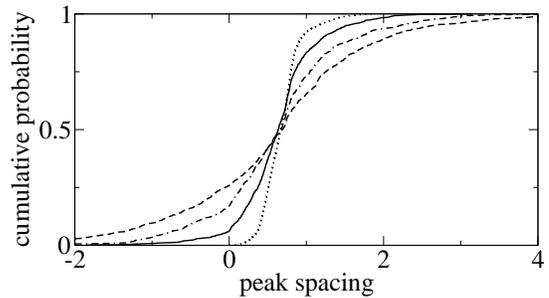}
\caption{Integrated  peak spacing  distribution  with $\zeta=0.8$  for
RMT/RPW  prediction   \cite{ullmo_baranger_01}  (dotted  line),  mostly
chaotic  (solid),  mixed   (dash-dot)  and  nearly  integrable  (dash)
regimes.
\label{fig:psp&spin}}
\end{figure}

The difference between the distribution of $M_{ij}$ observed for the
quartic oscillator system and that predicted for a chaotic system
is essentially that a non-negligible number of very localized
states $\psi_i$ give much larger diagonal terms, and much larger
variation of the corresponding off-diagonal terms.  The question is how
does this affect the ground state properties.  The answer for given
$(\lambda,\kappa)$ and number of electrons in the dot $N$ follows from
the $\e_i$ and $M_{ij}$.  To compute the energy $E[\{f_{i}^{\sigma}\}]$, we
use Eqs.~(\ref{eq:Egr},\ref{eq:E1p},\ref{eq:Eri}) for the various
occupations $\{f_{i}^{\sigma}\}$ such that $\sum_{i\sigma} f_{i}^{\sigma}
= N$.  The ground state follows by selecting the occupation sequence
minimizing this energy.  Varying $N$ in the range $[100,200]$ for each
parameter set, one constructs a distribution of total spins, occupancies,
and second differences in the ground state energy $\xi[N] = \Egr[N+1] +
\Egr[N-1] - 2 \Egr[N]$.  $\xi[N]$ is experimentally accessible by
measuring CB peak spacings, and we therefore refer to it as a ``peak
spacing''.   The various dynamical regimes' peak spacing distributions
are shown in Fig.~\ref{fig:psp&spin}.

We see  that the observed distributions are  strikingly different from
the RMT/RPW predictions (shown on the same plot) that should apply for
fully chaotic  systems.  Even our most chaotic  case shows significant
deviations, and  this increases considerably as the  system moves away
from  hard chaos.   The peak  spacings fluctuate  much more,  and very
large  spacings appear for  the mixed  and nearly  integrable regimes.
Moreover, as shown in Table~\ref{table:spinpol}, larger spins
become significantly more probable.
\begin{table}[b]
\begin{tabular}{|c|cccc|}
\hline
$\lambda$ & \ ~~+0.20~~ \ & \ ~~-0.20~~ \ & \  ~~-0.80~~ \ & ~RMT/RPW~
\\ \hline
$P(s=2)  $  & 0.13  &  0.16 & 0.07  & 0.01   \\
$~P(s=5/2)~$  & 0.08  &  0.10 & 0.02  & 0.00    \\
$\langle \delta s \rangle$ & 0.51  &  0.54 & 0.38   & 0.23  \\
\hline
\end{tabular}
\caption{\label{table:spinpol} Probabilities  $P(s=2)$ , $P(s=5/2)$ to
find a spin two (even $N$) or five halves (odd $N$) ground state, and
average value $\langle  \delta s  \rangle$ of the ground state spin
augmentation ($\delta s =  s$ or $(s-1/2)$ for even or odd number of
particles, respectively), for the  various dynamical regimes (values of
$\lambda$) with $\kappa=1.0$ and $\zeta=0.8$.  The last column is the
RMT/RPW prediction~\cite{ullmo_baranger_01}.}
\end{table}

\begin{figure}
\includegraphics[width=3.5in]{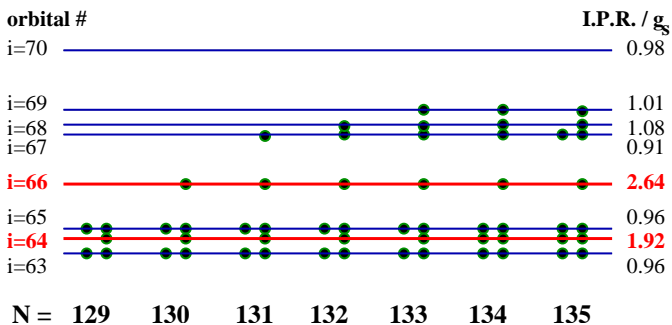}
\caption{Successive filling of the $(+,+)$ orbitals $i=63$ to $69$  as
the number of particles in the dot goes from $N=129$ to $135$ for
$(\lambda,\kappa)  =  (+0.20,1.00)$ and $\zeta=0.8$.   The  spacing
between the horizontal lines are proportional to the actual level
spacings.  The numbers in the right column are the corresponding
IPRs.
\label{fig:history}}
\end{figure}

We next ask how is such a drastic change made possible by the presence of
a moderate number of very localized states.  Let's consider one specific
example in detail.  Fig.~\ref{fig:history} shows the succession of
orbital fillings for a range of $N$.  Here, two of the single particle
states($i=64,66$), are highly localized.  Since the system is not chaotic,
there is also less level repulsion, which allows for the clustering
of levels around $\e_{64}$ or $\e_{69}$ that would otherwise be
improbable.
We see that none of the orbital occupations follow the simple ``up/down''
scenario characteristic of non-interacting systems.   As would seem
natural, the very localized orbitals $i=64,66$ remain singly occupied
across many values of $N$.  There is an additional non-intuitive feature,
namely that the other orbitals also prefer single occupancy despite not
being particularly localized.

This behavior derives from the following mechanism.  Suppose for
whatever reason, a given orbital $i^*$ is singly occupied, say with spin
$+$.  The $M_{ii*}$ have to be included in the total energy for {\em all}
orbitals $i$ if $(i,-)$ is occupied, but not if only $(i,+)$ is.  This can
make the energy cost of occupying an orbital with spin down $(\zeta
\Delta /g_s)$ higher than doing so with spin up.  Assuming $N_l$ such
singly occupied orbitals already exist, and adding the term $( \zeta
\Delta / g_s) M_{ii} $ associated with the interaction between the $(i,+)$
and the $(i,-)$ particles,  on average it turns out that (even for a
non-localized state) the residual interaction energy cost of doubly
occupying some orbital $i$ is $( (N_l+2) \zeta \Delta / g_s)$.  For typical
values of $\zeta$, this is larger than a mean spacing as soon as $N_l
\ge 1$.  Consequently, as illustrated by Fig~\ref{fig:history}, the
localized orbitals will not only remain singly occupied, but also have a
tendency to polarize the electrons in the other nearby orbitals.  Lack of
level repulsion and larger fluctuations of the $M_{ij}$ will further
enhance such effects.

To conclude, we have shown that for the non-TRI quartic oscillators
the eigenfunction statistics behave differently than those from a
RMT/RPW approach.  The significance is correlated with the degree of
chaos (or lack thereof) in the underlying classical dynamics.  Due to the
proximity of the Stoner instability, strong effects arise in ground state
spin polarizations, occupancies, and energies of the corresponding
``model'' quantum dot.   The quartic oscillators more fairly represent a
generic experimental dot than the hard chaos assumption.  The hard chaos
assumption leads to predictions that are qualitatively and quantitatively
incorrect.

Finally, such considerations should affect the understanding of realized
dots, which needs to be discussed on a case-by-case basis.  For instance,
the dots used in~\cite{ensslin} are certainly far from chaotic.  They
should show a large degree of phase space localization in their single
particle properties, whereas this point is debatable with respect
to the dots of~\cite{marcus_dot}.  However, it may be more interesting to
address this question from the opposite point of view.  Since creating
dots away from the hard chaos limit leads to behaviors that are
qualitatively different from those predicted using chaotic or diffusive
modeling, if the interest is in  devising a dot to perform some
particular function, such as spin manipulation, for instance, it is in the
soft chaos regime that richer behavior involving large fluctuations of
ground state energies and spins will be found.

We appreciate discussions with H.~Baranger and G. Usaj.  This work was
supported in part by NSF Grants \# PHY-0098027 and DMR-0103003.

\end{document}